\documentclass[twocolumn,letterpaper,nofootinbib,amsmath]{revtex4}

\pdfoutput=1
\usepackage{epsfig}
\usepackage{hyperref}

\newcommand{\lessthansimilarto}{\lower3pt\hbox{$\buildrel{<}\over{\sim}$}}
\newcommand{\greaterthansimilarto}{\lower3pt\hbox{$\buildrel{>}\over{\sim}$}}

\newcommand{\RR}{\hbox{$I$\kern-3.8pt $R$}}
\newcommand{\dperp}{\partial_\perp}

\begin{document}

%\title{Baumgarte--Shapiro--Shibata--Nakamura Formulation of General Relativity in Spherical Symmetry}
\title{BSSN in Spherical Symmetry}

\author{J.~David Brown}
\affiliation{Department of Physics, North Carolina State University,
Raleigh, NC 27695 USA}

\begin{abstract}
The BSSN (Baumgarte--Shapiro--Shibata--Nakamura) formulation of the Einstein 
evolution equations is written in spherical symmetry. These equations can be 
used to address a number of technical and conceptual issues in numerical 
relativity in the context of a single Schwarzschild black hole. One of the 
benefits of spherical symmetry is that the numerical grid points can be 
tracked on a Kruskal--Szekeres diagram. Boundary conditions suitable for 
puncture evolution of a Schwarzschild black hole are presented. Several 
results are shown for puncture evolution using a fourth--order finite 
difference implementation of the equations.  
\end{abstract}
 
\maketitle
%%%%%%%%%%%%%%%%%%
\section{Introduction}
The Baumgarte--Shapiro--Shibata--Nakamura \cite{Shibata:1995we,Baumgarte:1998te} formulation of the Einstein 
equations has been in widespread use in the numerical relativity 
community for a number of years. BSSN is the formulation used with the successful 
moving puncture technique for black hole 
simulations \cite{Campanelli:2005dd,Baker:2005vv}. Currently several groups are using BSSN with moving 
punctures to model the gravitational wave signal from binary black hole mergers.(See, for example, 
Refs.~\cite{Herrmann:2006ks,Sperhake:2006cy,Brugmann:2008zz,Gonzalez:2006md,Campanelli:2006fy,Baker:2006kr,Thornburg:2007hu,Tichy:2007hk}.)

There are still many open issues in numerical relativity, both technical and conceptual. 
One would like to explore these issues in an 
efficient manner. Current three--dimensional codes can take days or longer 
on a multi--processor computer cluster to carry out each 
simulation. By assuming spherical symmetry we can reduce the run time  to minutes  on 
a single processor. Thus there is an enormous practical advantage in assuming spherical symmetry. 
Of course, with a spherically symmetric (one--dimensional) code we are limited to issues that can be 
addressed in the context of a single Schwarzschild black hole. Furthermore there is no guarantee that the results 
one finds in spherical symmetry will transfer to the three--dimensional 
setting---in three dimensions there is a much wider range of possible problems that can arise. On the other hand, if a 
particular idea cannot be applied successfully in one dimension, it is unlikely to work in three. 

This article describes the one--dimensional code used in 
Refs.~\cite{Brown:2007tb,Brown:2007pg,turducken2} to explore 
the evolution of Schwarzschild black holes.
Such a code could also be used to examine various slicing and shift conditions, to explore different 
types of boundary conditions, and to experiment with constrained and unconstrained evolution schemes. 
Valuable insights into these and other problems can be gained with the help of a one--dimensional code. 

Many questions that one can address with a one--dimensional code 
are independent of the formulation of the Einstein evolution equations. 
But it is reasonable to expect that the behavior of a one--dimensional code based on the BSSN formulation will 
be closer to existing three--dimensional BSSN codes. This is important when  considering 
issues related to stability, constraint violations, {\it etc\/}. Also note that the $\Gamma$--driver 
shift condition, currently used for puncture evolution studies of binary black hole coalescence, makes 
explicit use of the conformal connection functions $\Gamma^a$. The conformal connection 
functions are basic variables in the BSSN formulation. 

There is a technical obstacle to writing the BSSN formulation of Einstein's equations in spherical symmetry. 
More generally, there is a technical obstacle to writing BSSN 
in spherical coordinates. The obstacle is this: The
BSSN equations assume that the conformal metric has determinant equal to one. This condition is not 
generally covariant. In  practice this is a problem because we typically use initial data that is conformally 
flat, and we naturally choose the initial conformal metric to be flat. But in spherical coordinates 
the flat metric does not have unit determinant. This obstacle can be overcome by generalizing the 
BSSN equations to allow for a conformal metric with non--unit determinant. The way to do so was 
presented in Ref.~\cite{Brown:2005aq}, and is reviewed in Sec.~2. 

The generalized BSSN equations  (GBSSN)  are invariant under conformal transformations \cite{Brown:2005aq}. 
This invariance is broken when one chooses an evolution equation for the determinant of the 
conformal metric. There are two natural choices that yield the ``Lagrangian case" and the ``Eulerian case". 
In Sec.~3 I show how  the GBSSN equations
are reduced to spherical symmetry for each of these cases. 
The level of hyperbolicity for these equations is discussed in Sec.~4. 

In this work I focus on puncture evolution for a Schwarzschild black hole. The puncture is located at the 
coordinate origin. In spherical coordinates, 
the origin is also a coordinate singularity. Clearly the boundary conditions at the origin play an important role. 
In Sec.~5 I describe sets of  boundary conditions that work well in practice. 
In Sec.~6 I show how the numerical grid points can be  mapped to a Kruskal diagram. 
This allows us to visualize the spacelike slice as it evolves, and track the distribution of grid points
relative to the horizon and physical singularity.
Section 7 presents some results for puncture evolution in one dimension. These results use 
the standard ``1+log'' slicing condition
\begin{equation}\label{onepluslog}
  \partial_t \alpha = \beta^a \partial_a\alpha - 2\alpha K \ ,
\end{equation}
or the related condition obtained by dropping the advection term $\beta^a\partial_a \alpha$. 
Here, $\alpha$ 
is the lapse function, $\beta^a$ is the shift vector, and $K$ is the trace of the extrinsic curvature. For the 
shift vector, the examples use either vanishing shift or the $\Gamma$--driver condition
\begin{subequations}\label{gammadriver}
\begin{eqnarray}
  \partial_t\beta^a & = & \frac{3}{4} B^a  + \beta^c\partial_c\beta^a \ ,\\
  \partial_t B^a & = & \partial_t {\Gamma}^a 
        + \beta^c\partial_c B^a - \beta^c\partial_c {\Gamma}^a -  \eta B^a \ .
\end{eqnarray}
\end{subequations}
Variants of the $\Gamma$--driver condition are obtained by dropping one or more of the 
advection terms $\beta^c\partial_c\beta^a$, $\beta^c\partial_c B^a$, or 
$\beta^c\partial_c\Gamma^a$ \cite{vanMeter:2006vi,Brugmann:2008zz}.

%%%%%%%%%%%%%%%%%%%%%%%%%%%%
\section{Generalized BSSN equations}
The BSSN equations can be generalized to allow for the possibility that the determinant of the conformal metric 
differs from unity \cite{Brown:2005aq}. The conformal metric ${ g}_{ab}$ and the trace--free 
part of the extrinsic curvature ${ A}_{ab}$ are defined by 
\begin{subequations}\label{physicalgandK}
\begin{eqnarray}
  \bar g_{ab} & = & e^{4\phi} { g}_{ab} \ ,\\
  K_{ab} & = & e^{4\phi} \left( { A}_{ab} + \frac{1}{3}{ g}_{ab} K \right) \ ,
\end{eqnarray}
\end{subequations}
where $\bar g_{ab}$ and $K_{ab}$ are the physical spatial metric and extrinsic curvature. The variable $\phi$ 
is the conformal factor and $K \equiv \bar g^{ab}K_{ab}$ is the trace of the physical extrinsic curvature.  The 
conformal connection functions are defined by 
\begin{equation}\label{Gammadef}
  {\Gamma}^a \equiv { g}^{bc} {\Gamma}^a_{bc} = -\frac{1}{\sqrt{ g}} 
     \partial_b \left( \sqrt{ g} { g}^{ab} \right) \ ,
\end{equation}
where ${\Gamma}^a_{bc}$ are the Christoffel symbols built from the conformal metric. The evolution 
equation for ${ \Gamma}^a$ is obtained by differentiating this definition and using the momentum 
constraint. The BSSN variables are $\phi$, ${ g}_{ab}$, ${ A}_{ab}$, $K$, and ${\Gamma}^a$. 

In vacuum, the generalized BSSN evolution equations (the GBSSN equations) 
are\footnote{Equation (\ref{bssnequations}c) corrects a sign error 
in Ref.~\cite{Brown:2005aq}.}
\begin{subequations}\label{bssnequations}
\begin{eqnarray}
  \partial_\perp \phi & = & -\frac{1}{12} \partial_\perp {\ln  g} -\frac{1}{6} \alpha K \ ,\\
  \partial_\perp { g}_{ab} & = & \frac{1}{3} { g}_{ab} \partial_\perp{\ln  g} 
      - 2\alpha { A}_{ab} \ ,\\
  \partial_\perp{ A}_{ab} & = & \frac{1}{3} { A}_{ab} \partial_\perp{\ln  g}
      - 2\alpha { A}_{ac}{ A}^c_b + \alpha { A}_{ab} K  \nonumber\\
      & &\!\! +\, e^{-4\phi}\left[ \alpha \bar R_{ab} - \bar D_a \bar D_b\alpha \right]^{\rm TF} \ ,\\
  \partial_\perp K & = & \frac{1}{3} \alpha K^2 + \alpha { A}_{ab}{ A}^{ab} 
	- {\bar D}^a {\bar D}_a \alpha \ ,\\
  \partial_\perp { \Gamma}^a & = & -\frac{1}{3} { \Gamma}^a \partial_\perp {\ln g} 
      - \frac{1}{6} { g}^{ab}\partial_b \partial_\perp{\ln g} 
      - 2{ A}^{ab}\partial_b \alpha \nonumber\\ 
      & & \!\! +\, 2\alpha \left[ {\Gamma}^a_{bc}{ A}^{bc} 
      + 6{ A}^{ab}\partial_b \phi - \frac{2}{3} { g}^{ab}\partial_b K \right] \ ,
\end{eqnarray}
\end{subequations}
where $\bar R_{ab}$ is the physical Ricci tensor, $\bar D_a$ is the physical covariant derivative,  
and ${ D}_a$ is the conformal covariant derivative. 
The shift vector $\beta^a$ is buried in the time derivative operator, 
$\partial_\perp \equiv \partial_t - {\cal L}_\beta$.  Also note that TF stands for the trace--free part of the 
expression in brackets. 

The  variables $\phi$, ${ g}_{ab}$, 
${ A}_{ab}$, and $K$ are defined as tensors with no density weights. 
The transformation rules for the conformal connection functions 
${\Gamma}^a$ are determined by the transformation rules for the Christoffel symbols ${\Gamma}^a_{bc}$ 
and the definition Eq.~(\ref{Gammadef}). 
The GBSSN equations do not assume $ g = 1$, and 
we are free to choose how $ g$  evolves in time. Two natural choices are the 
``Eulerian condition" $\partial_\perp\ln g = 0$ and the ``Lagrangian condition" 
$\partial{(\ln g)}/\partial t = 0$, which implies  $\partial_\perp {\ln g} = 
-2{ D}_a\beta^a$. 

In the traditional BSSN formulation, the restriction $g=1$ appears as an extra constraint. 
In most numerical codes this constraint is enforced by replacing $g_{ab}$ with $g_{ab}/g^{1/3}$ at the 
end of each time step. There is no clear equivalence between the traditional BSSN formulation and 
GBSSN with either the Eulerian (GBSSN-E) or Lagrangian (GBSSN-L) conditions. In particular, observe 
that there are $16$ field 
variables evolved by the GBSSN equations (namely $\phi$, the six components of $g_{ab}$, the 
five components of $A_{ab}$, $K$, and the three components of $\Gamma^a$) but only $15$ 
variables in the traditional BSSN system. One consequence of this difference is a mismatch 
in the number of characteristic fields for GBSSN and traditional BSSN. 
In spite of this difference one might guess that GBSSN with the 
Lagrangian condition $\partial g/\partial t = 0$  is closely related to 
traditional BSSN. With GBSSN-L the determinant $g$, which can be set to $1$ by initial conditions, 
remains constant throughout the evolution. Indeed, a qualitative comparison between the results 
presented in Section VII and results obtained in three dimensions with traditional BSSN 
shows that GBSSN-L and traditional BSSN yield very similar 
behavior for puncture evolution of a single Schwarzschild black hole. For example, compare 
Figs.~7--10 of this paper with Fig.~1 of Ref.~\cite{Brugmann:2008zz}. 

Although GBSSN-L and traditional BSSN are similar in some respects, GBSSN-L 
contains an extra characteristic field. That field travels along the integral curves of the 
time flow vector field $\partial_t$. Thus, this field has a characteristic speed 
that, when measured with respect to observers at rest in the 
spacelike hypersurfaces, depends on the shift vector. This is not necessarily bad, but it is at 
least unusual and unphysical. For a typical formulation of the Einstein equations, the characteristic 
speeds are independent of the shift vector as long as any dynamical gauge conditions are expressed 
in terms of the normal derivative operator $\dperp = \partial_t - {\cal L}_\beta$. GBSSN-L breaks 
this pattern. On the other hand, for GBSSN-E, the 
extra characteristic is a zero--speed mode. As a result, the Eulerian condition yields a more simple 
and in some sense more physical version of the GBSSN system. 

Finally, note that the variable $\phi$ can be replaced with a new variable 
$\chi \equiv e^{-4\phi}$. It is straightforward to 
rewrite the GBSSN equations in terms of $\chi$. Note that $\chi$, like $\phi$, is a scalar 
with no density weight. The variable $\phi$ was used in Ref.~\cite{Baker:2005vv}, while the authors of 
Ref.~\cite{Campanelli:2005dd} used $\chi$. Reference~\cite{Brugmann:2008zz} includes comparisons 
between the two choices for black hole puncture evolution.

%%%%%%%%%%%%%%%%%%%%%%%%%%%%%
\section{GBSSN reduced by spherical symmetry}
Let the spatial coordinates be denoted by $r$, $\theta$, and $\varphi$. 
The GBSSN equations are 
reduced to spherical symmetry with the following ansatz for the metric:
\begin{equation}\label{g_ss}
  { g}_{ab} = \left( \begin{array}{ccc} { g}_{rr} & 0 & 0 \\
                                         0 & { g}_{\theta\theta} & 0 \\
                                         0 & 0 & { g}_{\theta\theta}\sin^2{\theta} 
                            \end{array}  \right)
\end{equation}
The ansatz for the trace--free part of the extrinsic curvature is
\begin{equation}\label{A_ss}
  { A}_{ab} = { A}_{rr} \left( \begin{array}{ccc} 1 & 0 & 0 \\
                      0 & -{ g}_{\theta\theta}/({2{ g}_{rr}}) & 0 \\
		      0 & 0 & -{ g}_{\theta\theta}\sin^2{\theta}/({2{ g}_{rr}})
		      \end{array} \right)
\end{equation}
and for the conformal connection functions,
\begin{equation}\label{Gamma_ss}
  { \Gamma}^a = \left( \begin{array}{c} {\Gamma}^r \\
                         -{\cos{\theta}}/({g_{\theta\theta}\sin{\theta}}) \\
			 0 \end{array} \right)
\end{equation}
The dynamical variables are now $\phi$ or $\chi$, ${ g}_{rr}$, 
${ g}_{\theta\theta}$, ${ A}_{rr}$, $K$, 
and ${\Gamma}^r$. 
They are all functions of the spatial coordinate $r$ and time $t$. 

For the rest of this paper I will use the variable $\chi$ rather than $\phi$. I also 
use $\partial_\perp{\ln g} = -2v{ D}_a\beta^a$ where $v=0$ gives the Eulerian condition 
and $v=1$ gives the Lagrangian condition. 
In spherical symmetry, the GBSSN equations are as follows:
\vfill\eject
\begin{widetext}
\begin{subequations}
\begin{eqnarray}
	\partial_t\chi & = & \frac{2 K \alpha  \chi }{3}-\frac{v \beta^r  {g_{rr}}' \chi }{3 {g_{rr}}}-\frac{2 v \beta^r  {g_{\theta\theta}}' \chi }{3 {g_{\theta\theta}}}-\frac{2}{3} v {\beta^r}' \chi +\beta^r  \chi ' 
   \ ,\label{chieqn}\\
   \partial_t { g}_{rr} & = & -2 {A_{rr}} \alpha -\frac{1}{3} v \beta^r  {g_{rr}}'+\beta^r  {g_{rr}}'-\frac{2 {g_{rr}} v \beta^r  {g_{\theta\theta}}'}{3 {g_{\theta\theta}}}+2 {g_{rr}} {\beta^r}'-\frac{2}{3} {g_{rr}} v
   {\beta^r}' 
  \ ,\label{grreqn}\\
  \partial_t { g}_{\theta\theta} & = & \frac{{A_{rr}} {g_{\theta\theta}} \alpha }{{g_{rr}}}-\frac{{g_{\theta\theta}} v \beta^r  {g_{rr}}'}{3 {g_{rr}}}-\frac{2}{3} v \beta^r  {g_{\theta\theta}}'+\beta^r  {g_{\theta\theta}}'-\frac{2}{3}
   {g_{\theta\theta}} v {\beta^r}' 
  \ , \label{gthetathetaeqn}\\
  \partial_t  A_{rr} & = & -\frac{2 \alpha  {A_{rr}}^2}{{g_{rr}}}+K \alpha  {A_{rr}}-\frac{v \beta^r  {g_{rr}}' {A_{rr}}}{3 {g_{rr}}}-\frac{2 v \beta^r  {g_{\theta\theta}}' {A_{rr}}}{3
   {g_{\theta\theta}}}-\frac{2}{3} v {\beta^r}' {A_{rr}}+2 {\beta^r}' {A_{rr}}+\frac{2 \alpha  \chi  \left({g_{rr}}'\right)^2}{3 {g_{rr}}^2}    \nonumber \\  & & -\frac{\alpha  \chi 
   \left({g_{\theta\theta}}'\right)^2}{3 {g_{\theta\theta}}^2}-\frac{\alpha  \left(\chi '\right)^2}{6 \chi }   -\frac{2 {g_{rr}}
   \alpha  \chi }{3 {g_{\theta\theta}}}+\beta^r  {A_{rr}}'+\frac{2}{3} {g_{rr}} \alpha  \chi  {\Gamma^r}'  -\frac{\alpha  \chi  {g_{rr}}' {g_{\theta\theta}}'}{2 {g_{rr}} {g_{\theta\theta}}}+\frac{\chi  {g_{rr}}' \alpha '}{3 {g_{rr}}}\nonumber \\ & &  +\frac{\chi  {g_{\theta\theta}}' \alpha '}{3
   {g_{\theta\theta}}}-\frac{\alpha  {g_{rr}}' \chi '}{6 {g_{rr}}}-\frac{\alpha  {g_{\theta\theta}}' \chi '}{6 {g_{\theta\theta}}}-\frac{2 \alpha ' \chi '}{3}-\frac{\alpha  \chi 
   {g_{rr}}''}{3 {g_{rr}}}+\frac{\alpha  \chi  {g_{\theta\theta}}''}{3 {g_{\theta\theta}}}-\frac{2 \chi  \alpha ''}{3}+\frac{\alpha  \chi ''}{3} 
  \ , \label{Arreqn}\\
   \partial_t K & = & \frac{3 \alpha  {A_{rr}}^2}{2 {g_{rr}}^2}+\frac{K^2 \alpha }{3}+\beta^r  K'+\frac{\chi  {g_{rr}}' \alpha '}{2 {g_{rr}}^2}-\frac{\chi  {g_{\theta\theta}}' \alpha '}{{g_{rr}}
   {g_{\theta\theta}}}+\frac{\alpha ' \chi '}{2 {g_{rr}}}-\frac{\chi  \alpha ''}{{g_{rr}}} 
  \ , \label{Keqn}\\
  \partial_t \Gamma^r & = & -\frac{v \beta^r  \left({g_{\theta\theta}}'\right)^2}{{g_{rr}} {g_{\theta\theta}}^2}+\frac{{A_{rr}} \alpha  {g_{\theta\theta}}'}{{g_{rr}}^2 {g_{\theta\theta}}}-\frac{v {\beta^r}' {g_{\theta\theta}}'}{3 {g_{rr}} {g_{\theta\theta}}}+\frac{{\beta^r}' {g_{\theta\theta}}'}{{g_{rr}} {g_{\theta\theta}}}+\beta^r  {\Gamma^r}' +\frac{{A_{rr}} \alpha  {g_{rr}}'}{{g_{rr}}^3}-\frac{4 \alpha  K'}{3 {g_{rr}}}-\frac{2 {A_{rr}} \alpha '}{{g_{rr}}^2}  \nonumber\\  & &  +\frac{v
   {g_{rr}}' {\beta^r}'}{2 {g_{rr}}^2}-\frac{{g_{rr}}' {\beta^r}'}{2 {g_{rr}}^2}-\frac{3 {A_{rr}} \alpha  \chi '}{{g_{rr}}^2 \chi }+\frac{v \beta^r  {g_{rr}}''}{6
   {g_{rr}}^2}+\frac{v \beta^r  {g_{\theta\theta}}''}{3 {g_{rr}} {g_{\theta\theta}}}+\frac{v {\beta^r}''}{3 {g_{rr}}}+\frac{{\beta^r}''}{{g_{rr}}} 
  \ ,\label{Gammaeqn}
\end{eqnarray}
\end{subequations}
Primes denote derivatives with respect to $r$. 
Note that I am following the common practice of using $\Gamma^r$ on the right--hand side only if it 
appears differentiated. 

The Hamiltonian constraint is defined by ${\cal H} \equiv K^2 - K_{ab}K^{ab} + \bar R$ 
and the momentum constraint is defined by ${\cal M}_a \equiv \bar D_b K^b_a - \bar D_aK$. (Indices 
on $K_{ab}$ are raised with the physical metric.) With BSSN we also have 
constraints that arise from the definition of the conformal connection functions: ${\cal G}^a \equiv {\Gamma}^a 
- { g}^{bc} {\Gamma}_{bc}^a$. In spherical symmetry these constraints become
\begin{subequations}\label{constraints}
\begin{eqnarray}
  {\cal H} & = & -\frac{3 { A_{rr}}{}^2}{2 {{ g}_{rr}}{}^2}+\frac{2 K^2}{3}-\frac{5
   {\chi'}{}^2}{2 \chi  {{ g}_{rr}}}+\frac{2
   {\chi''}}{{{ g}_{rr}}}+\frac{2 \chi }{{{ g}_{\theta\theta}}}-\frac{2 \chi 
   {{ g_{\theta\theta}}''}}{{{ g}_{rr}} {{ g}_{\theta\theta}}}+\frac{2 {\chi'}
   {{ g_{\theta\theta}}'}}{{{ g}_{rr}} {{ g}_{\theta\theta}}}+\frac{\chi  {{ g_{rr}}'}
   {{ g_{\theta\theta}}'}}{{{ g}_{rr}}{}^2 {{ g}_{\theta\theta}}}-\frac{{\chi'}
   {{ g_{rr}}'}}{{{ g}_{rr}}{}^2}+\frac{\chi  { {g_{\theta\theta}}'}{}^2}{2 {{ g}_{rr}}
   {{ g}_{\theta\theta}}{}^2}
  \ ,\label{Heqn} \\
  {\cal M}_r & = &  \frac{ { {A_{rr}}'}}{{{ g}_{rr}}}-\frac{2  {K'}}{3}-\frac{3
   { A_{rr}} {\chi'}}{2 \chi {{ g}_{rr}}}+\frac{3 { A_{rr}} 
   {{ g_{\theta\theta}}'}}{2 {{ g}_{rr}} {{ g}_{\theta\theta}}}-\frac{{ A_{rr}} 
   {{ g_{rr}}'}}{{{ g}_{rr}}{}^2}
  \ ,\label{Meqn} \\
  {\cal G}^r & = & -\frac{{{ g_{rr}}'}}{2
   { g}_{rr}{}^2}+{\Gamma^r}+\frac{{ g_{\theta\theta}}'}{{ g}_{rr}
   { g}_{\theta\theta}} \ ,\label{Geqn} 
\end{eqnarray}
\end{subequations}
and the constraint evolution system is 
\begin{subequations}\label{conevsystem}
\begin{eqnarray}
           \partial_t{\cal H} & = & \beta^r {\cal H}' + \frac{2}{3}\alpha K {\cal H}
                - \frac{2\alpha A_{rr} \chi}{g_{rr}} {{\cal G}^r}'
               -\frac{2\alpha\chi}{g_{rr}}{{\cal M}_r}' + \left[ \frac{\alpha\chi'}{g_{rr}} 
                + \frac{\alpha\chi {g_{rr}}'}{{g_{rr}}^2} - \frac{4\alpha'\chi}{g_{rr}} 
                - \frac{2\alpha\chi {g_{\theta\theta}}'}{g_{rr}g_{\theta\theta}} \right] {\cal M}_r
                    \ ,\\
        \partial_t{\cal M}_r & = & \beta^r {\cal M}_r{}' + {\beta^r}'{\cal M}_r 
                + \alpha K {\cal M}_r - \frac{\alpha'}{3} {\cal H} 
                + \frac{\alpha}{6} {\cal H}' + \frac{2}{3} \alpha\chi{{\cal G}^r}'' 
	        + \left[ \frac{2\alpha'\chi}{3} - \frac{\alpha\chi'}{3} 
                    + \frac{\alpha\chi {g_{\theta\theta}}'}{g_{\theta\theta}} \right] {{\cal G}^r}'
                              \ ,\\
        \partial_t{\cal G}^r & = & \beta^r {{\cal G}^r}' + \frac{2\alpha}{g_{rr}} {\cal M}_r \ .
\end{eqnarray}
\end{subequations}
\end{widetext}

In the three--dimensional application of BSSN one must face the possibility that under numerical 
evolution ${ A}_{ab}$ will not remain trace--free. In most three--dimensional codes the 
components ${ A}_{ab}$ are adjusted at the end of each timestep to maintain the constraint 
${ A}_{ab} { g}^{ab} =0$. 
In spherical symmetry ${ A}_{ab}$ is automatically traceless by virtue of the ansatz Eq.~(\ref{A_ss}). 
As discussed previously, the determinant of the conformal metric is not constrained in the generalized 
BSSN formulation. 

%%%%%%%%%%%%%%%%%%%%%%%%%%%%%%%%%%%
\section{Hyperbolicity}
Consider the Eulerian ($\partial_\perp g = 0$) and Lagrangian ($\partial_\perp g = -2D_a\beta^a$) 
formulations with 1+log slicing Eq.~(\ref{onepluslog}) and either $\Gamma$--driver shift or 
vanishing shift. With the 1+log slicing and $\Gamma$--driver shift conditions we can either 
include or exclude the advection terms. These cases will be labeled $E$ for Eulerian and $L$ for 
Lagrangian. 
A superscript $+$ or $-$ will denote the inclusion or exclusion of the advection term in 
the 1+log slicing condition. A subscript $+$ or $-$ will denote the inclusion or exclusion 
of all advection terms in the $\Gamma$--driver shift condition. For example, 
$E_+^+$ represents the Eulerian case with 1+log slicing that includes the advection term 
and $\Gamma$--driver shift that includes all advection terms. 
Similarly, $L^+_-$ denotes the Lagrangian case with 1+log slicing that includes the 
advection term and $\Gamma$--driver shift that excludes all
advection terms. I also use a subscript 
$0$ to denote 
vanishing shift. In these cases the advection term in the 1+log slicing condition vanishes, 
so the superscript can be omitted. 
Thus,  $E_0$ and $L_0$ refer to the Eulerian and Lagrangian cases with vanishing shift. 

The results on hyperbolicity reported here were obtained using pseudo--differential 
techniques \cite{Nagy:2004td,Beyer:2004sv}. These techniques apply to systems of PDE's that have first order time
derivatives and second order space derivatives. Pseudo--differential methods can only distinguish 
between weak and strong hyperbolicity; other methods must be used to check for symmetric hyperbolicity. 

For  the cases $E_0$, $L_0$, which have vanishing shift, the full system (equations of motion and 1+log slicing) is 
strongly hyperbolic. For the cases with nonvanishing shift,  the full system (equations of 
motion plus gauge conditions) is strongly hyperbolic as long as certain inequalities among 
the field variables are satisfied. 

There are three characteristic fields that are common to all cases. These are: 
\begin{subequations}\label{charfields}
\begin{eqnarray}
   & & \Gamma^r \mp \frac{3}{2\sqrt{g_{rr}^3\chi}} A_{rr} + \frac{1}{2g_{rr}\chi} \chi'  
     - \frac{1}{2g_{rr}^2} g_{rr}{}'  \nonumber \\ & & \qquad   + \frac{1}{2g_{rr}g_{\theta\theta}} g_{\theta\theta}{}'  
    \pm \frac{1}{\sqrt{g_{rr}\chi}} K  \ ,\\
   & & \Gamma^r + \frac{2}{\chi g_{rr}} \chi' - \frac{1}{2 g_{rr}^2} g_{rr}{}' 
     - \frac{1}{g_{rr} g_{\theta\theta}} g_{\theta\theta}{}' \ .
\end{eqnarray}
\end{subequations}
The fields (\ref{charfields}a) have proper speeds $\pm 1$ as measured by observers at rest in the spacelike slices. 
The field (\ref{charfields}b) has vanishing speed. All cases in which the advection term is included in 
the 1+log slicing condition (or the shift vanishes) have characteristic fields
\begin{equation}\label{charfield2}
	\alpha' \pm \sqrt{2\alpha g_{rr}/\chi} K 
\end{equation} 
with proper speeds $\pm \sqrt{2/\alpha}$. 

The remaining characteristic fields and speeds depend on the details of the formulation and the gauge 
conditons. For  $E_0$ there are two characteristic fields in addition to those 
displayed in Eqs.~(\ref{charfields}), (\ref{charfield2}). Both are zero speed modes. 
For $L_0$, the 
remaining characteristic fields have speeds $0$ and $\hat\beta^r$, where 
$\hat\beta^r \equiv \sqrt{g_{rr}/\chi} \beta^r/\alpha$ is the proper length shift per unit proper time. 

For the cases $E^+_\pm$ and $L^+_\pm$ there are four characteristic fields 
in addition to those displayed above. For $E^+_+$ the extra characteristic fields have speeds  
$0$, $0$, $\pm \sqrt{3/(\alpha^2\chi)}/2$. This system is strongly hyperbolic as long 
as $8\alpha \chi \ne 3$. For $L^+_+$, the remaining characteristic fields 
have speeds $0$, $\hat\beta^r$, $\pm\sqrt{1/(\alpha^2\chi)}$ and the requirements for 
strong hyperbolicity are  $\sqrt{g_{rr}}|\beta^r|  \ne 1$ and $2\alpha\chi \ne 1$. 
For the case $E^+_-$, the remaining characteristic fields 
have speeds $0$, $\hat\beta^r$, $\Bigl[\hat\beta^r \pm \sqrt{(\hat\beta^r)^2 + 3/(\alpha^2\chi)}\Bigr]/2$ and the 
requirement for strong hyperbolicity is $(8\alpha\chi - 3)  \ne \pm 4 \sqrt{2\alpha\chi g_{rr}} \beta^r$.  
For the case $L^+_-$, the remaining characteristic 
fields have speeds $\hat\beta^r$, 
$\hat\beta^r$, $\Bigl[\hat\beta^r \pm \sqrt{(\hat\beta^r)^2 + 4/(\alpha^2\chi)}\Bigr]/2$ 
and the requirement for strong hyperbolicity is 
$2\alpha\chi - 1 \ne \pm \sqrt{2\alpha\chi g_{rr}} \beta^r$. 
In all cases, strong hyperbolicity can be spoiled if one or more of the fields 
$g_{rr}$, $g_{\theta\theta}$, $\chi$ or $\alpha$ vanishes. I have not carried out a complete analysis 
of the restrictions on hyperbolicity for the cases $E^-_\pm$ and $L^-_\pm$. 

The restrictions on strong hyperbolicity are not a consequence of the reduction to spherical symmetry. 
Precisely the same restrictions are found for the full three--dimensional GBSSN systems. In three dimensions 
one analyzes hyperbolicity by choosing a unit normal covector $n_a$ and projecting the principal
parts of the equations of motion in directions normal and tangential to $n_a$. In this way the principal
symbol splits into blocks that transform as scalars, vectors, and trace--free tensors under rotations about the 
normal direction. The vector and tensor blocks are strongly hyperbolic without restriction. The scalar 
block in three dimensions is identical to the principal symbol for the spherically symmetric GBSSN system. 
It follows that 
GBSSN has the same restrictions on strong hyperbolicity in three dimensions as it has in spherical
symmetry. 

It should be noted that for traditional BSSN with 1+log slicing and $\Gamma$--driver shift, including 
all advection terms, strong hyperbolicity requires $2\alpha\chi  \ne 1$ \cite{Beyer:2004sv}. 
For traditional BSSN with 1+log slicing and $\Gamma$--driver shift and  
various combinations of advection terms, the restrictions on hyperbolicity have been 
analyzed in Ref.~\cite{Gundlach:2006tw}. 

It is worth noting that the speeds for some of the characteristic fields can exceed the speed of light. 
For example, the system $E_+^+$ has a mode with speed $\sqrt{2/\alpha}$ that is superluminal 
for $\alpha < 2$, and a mode with speed 
$\sqrt{3/(\alpha^2\chi)}/2$ that is superluminal for $\alpha^2\chi < 3/4$. These modes are 
associated with  the 1+log and $\Gamma$--driver gauge conditions \cite{Garfinkle:2007yt}. As such, they 
appear in any formulation of the Einstein evolution equations 
that uses 1+log slicing and $\Gamma$--driver shift.  

The constraint evolution system Eq.~(\ref{conevsystem}) is strongly hyperbolic with characteristic speeds 
$0$, $\pm 1$. The respective characteristic fields are 
\begin{subequations}
\begin{eqnarray}
   & & {\cal H} + \chi {\cal G}^r \ , \\
  & & {\cal H} \mp 6\sqrt{\chi/g_{rr}} {\cal M}_r + 4\chi {\cal G}^r{}'   \ .
\end{eqnarray}
\end{subequations}
In three dimensions,  the constraint propagation systems for both the GBSSN equations and the 
traditional BSSN equations are strongly hyperbolic with causal characteristic 
speeds $0$ and $\pm 1$. 

%%%%%%%%%%%%%%%%%%%%%%%%%%%%%%%%%%
\section{Boundary conditions}
The one--dimensional code described in this paper could be used to evolve a single black hole 
with excision, or to evolve smooth spherically symmetric fields in $\RR^3$, provided appropriate 
boundary conditions are imposed. For the case of fields that are smooth at the origin, such as 
the fields that describe a spherically symmetric 
star, boundary conditions have been discussed in detail 
in Refs.~\cite{Alcubierre:2004gn,Ruiz:2007rs} and elsewhere. 
In this  paper I will focus on boundary conditions that 
allow for black hole puncture evolution. 
In this case the origin $r=0$ defines the puncture and is also a coordinate singularity in the underlying 
spherical coordinate system. The grid is cell centered with spacing $\Delta r$. The grid points 
are located at coordinate radii $r(j) = (j - 1/2)\Delta r$, where $j = 1,2,\ldots$. 

In applications involving smooth fields with spherical coordinates one would normally 
impose smoothness conditions at the origin \cite{Alcubierre:2004gn,Ruiz:2007rs}. 
Specifically, scalars and type 2 tensors would be reflection symmetric and vectors would be 
antisymmetric. With puncture evolution we need to allow for the possibility that the fields will not remain smooth, 
even if they are smooth initially. We can nevertheless attempt to 
impose smoothness conditions at the puncture. In my experience this leads to an unstable code. 

To obtain appropriate boundary conditions  for puncture evolution 
I take a minimalist approach. That is, my goal is to  impose as little information as possible through boundary 
conditions so that  the puncture can evolve without unnecessary influence from the outside. After some experimentation, 
I  found that the following prescription works well for the five Lagrangian cases $L_0$ and $L^\pm_\pm$. 
For the 
variables ${ g}_{rr}$, $\chi$, ${ A}_{rr}$, $K$, ${\Gamma}^r$, $\alpha$, and $B^r$, no boundary 
conditions are imposed. That is, for grid points near the puncture, the finite difference stencil is shifted away 
from the puncture so that no guard cells are needed. 

Typically, the condition I impose on ${ g}_{\theta\theta}$ and $\beta^r$ is that these variables should
vanish at the puncture. In keeping with the minimalist approach I only use one layer of guard cells. 
The guard cells ($j=0$) are filled  via the relations 
\begin{subequations}\label{LagrangianBC}
\begin{eqnarray}
  { g}_{\theta\theta}(0) & = &  [-315{ g}_{\theta\theta}(1) + 210{ g}_{\theta\theta}(2) 
  - 126{ g}_{\theta\theta}(3) \nonumber\\
  & & + 45g_{\theta\theta}(4) - 7g_{\theta\theta}(5) ]/63 \ , \\
  \beta^r(0) & = &  [-315\beta^r(1) + 210\beta^r(2) 
  - 126\beta^r(3) \nonumber\\
  & & + 45 \beta^r(4) - 7\beta^r(5) ]/63  \ ,
\end{eqnarray}
\end{subequations}
where the numbers in parentheses denote grid points. These relations insure that ${ g}_{\theta\theta}$ 
and $\beta^r $ vanish at the puncture to sixth order accuracy. In other words, a fifth order polynomial fit 
across the grid points $j=0\ldots 4$ with value (\ref{LagrangianBC}) at $j=0$ will yield a function that 
vanishes at $r=0$. 

The results presented in Sec.~VII are obtained from  a code that uses 
fourth order accurate spatial differencing. In the bulk of the computational domain, I use centered 
stencils that extend across five grid points for first and second spatial derivatives. (An exception 
to this rule is made for advection terms like $\beta^r\partial_r {g}_{rr}$. For these terms, 
the finite difference stencil is shifted by one grid point in the ``upwind'' direction.) Since only one guard
cell is available at the boundary $r=0$, the stencils near $r=0$ are shifted so that they require just one 
grid point on the side closest to the puncture. 

For the Eulerian cases $E_0$, $E^+_+$ and $E^-_+$, no boundary conditions are imposed on the variables ${ g}_{rr}$, 
$g_{\theta\theta}$, $\chi$, ${ A}_{rr}$, $K$, ${\Gamma}^r$, $\alpha$, and $B^r$. The only variable that remains 
is the shift vector $\beta^r$. For the shift I impose $\partial_r\beta^r = 0$ via the relationship 
\begin{equation}\label{EulerianBC}
   \beta^r(0) = [17\beta^r(1) + 9\beta^r(2) - 5\beta^r(3) + \beta^r(4)]/22 \ .
\end{equation}
Here again the numbers in parentheses label grid points. A fourth order polynomial fit across the 
grid points $j=0\ldots 4$, with value (\ref{EulerianBC}) at $j=0$, will yield a function with vanishing 
derivative at $r=0$. 

For the cases $E^+_-$ and $E^-_-$ with the condition Eq.~(\ref{EulerianBC}), an instability 
develops at the puncture within a time of a few $M$. 
I have not found suitable inner boundary conditions for these cases. 

\begin{figure}[t!]
\includegraphics[scale=0.9, viewport=135 350 440 540]{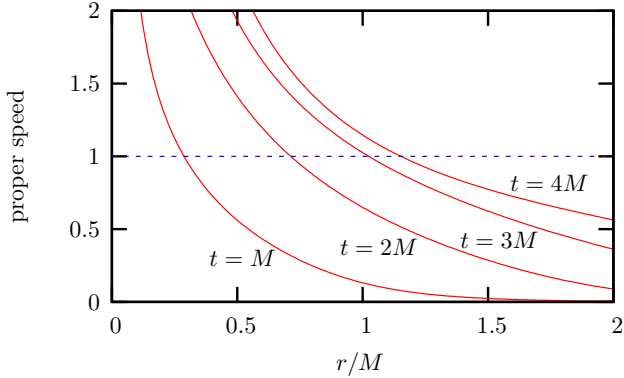}
\caption{The proper speed $\sqrt{(\beta^r)^2 g_{rr}/\chi}/\alpha$ of the coordinate system, 
plotted as a function of coordinate radius $r$ at times $t = M$, $2M$, $3M$, and $4M$.}
\label{figure2speed}
\end{figure}

The minimalist approach to boundary conditions at $r=0$ is not unreasonable. Simulations show that 
as the data evolve from the initial conditions, the light cones at grid points near the puncture quickly tip outward (toward $r=0$) with respect to the time flow 
vector field. In Fig.~(\ref{figure2speed}) the proper speed of the coordinate system  with respect to the 
spacelike slices is plotted at various times up to $t=4M$. During the first $4M$ of evolution the horizon 
moves from $r=M/2$ to just beyond $r=M$.
From the graph we see that by $t=M$ all of the grid points inside $r\approx M/3$ are moving faster 
than the speed of light (which is unity). 
By $t=4M$ the region of superluminal grid movement has expanded to about $r\approx M$, the location of the horizon. 
There is little change beyond $t=4M$.  

Once the time flow direction moves outside the light cone, one might expect that any boundary conditions 
imposed at the puncture would be irrelevant. This is not quite correct because GBSSN (as well as 
traditional BSSN) with 1+log slicing and $\Gamma$--driver shift 
contains characteristic modes that can travel faster than light. For example, 
the $E_+^+$ system has modes with speeds $\pm\sqrt{2/\alpha}$ and $\pm\sqrt{3/(4\alpha^2\chi)}$.
The $L_+^+$ system has modes with speeds $\pm\sqrt{2/\alpha}$, $\pm\sqrt{1/(\alpha^2\chi)}$, 
and $\hat\beta^r$. 

Ideally, one should impose boundary conditions on all incoming 
characteristic fields at the puncture. Whether or not a characteristic field is incoming or outgoing 
depends on the proper speed of the mode relative to the proper speed of the coordinate grid, which is 
$\hat\beta^r$. For a typical puncture evolution the modes listed in Eq.~(13) with 
speeds $\pm\sqrt{2/\alpha}$ are outgoing, at least for times beyond the first few $M$. 
For $E_+^+$, the characteristic field 
\begin{equation}\label{coordmode}
	{\beta^r}' - \frac{\sqrt{3g_{rr}}}{2} B^r + \frac{8\alpha\chi}{\sqrt{3g_{rr}} 
	(8\alpha\chi - 3)} \alpha' + \frac{4\alpha}{8\alpha\chi - 3} K
\end{equation}
has speed $\pm\sqrt{3/(4\alpha^2\chi)}$ and is incoming at the puncture 
throughout the evolution. Thus, we would expect that certain boundary information must be 
imposed at the puncture to fix this incoming mode. The condition 
Eq.~(\ref{EulerianBC}) is not ideal, but appears to be sufficient to allow for stable evolutions. 
The case $L_+^+$ is more difficult to analyze, because in addition to an incoming mode it also contains 
a mode that travels along the puncture boundary. 

The above discussion shows that for puncture evolution in one dimension, some 
boundary conditions at the puncture are necessary. This might come as a surprise since, apparently, 
no boundary conditions are imposed at the puncture in three--dimensional codes. 
This is probably an illusion. More likely, the incoming modes in the three--dimensional 
case are fixed implicitly through the numerical (finite differencing) 
scheme used in the vicinity of the puncture. 

The details of the finite differencing scheme in the vicinity of the puncture 
would have little affect on the data away from the puncture, and would be especially 
difficult to detect outside the black hole. 
Evidence for this assertion can be found in recent work on the 
``turducken" approach to black hole evolution \cite{Etienne:2007hr,Brown:2007pg,turducken2}, 
in which the black hole interior is stuffed with artificial, constraint--violating data.
In this case it is found, both for one--dimensional simulations with $E_+^+$ and for three--dimensional 
simulations with traditional BSSN, that the stuffing can only affect the slicing and coordinate  
conditions outside the black hole \cite{turducken2}. The slicing can be changed outside the black hole 
due to the presence of the mode  Eq.~(13) whose speed $\sqrt{2/\alpha}$ can become superluminal. 
However, inside a coordinate radius of about $r\sim 0.2M$, this mode moves more slowly than the 
coordinate grid and cannot propagate to the black hole exterior. 
The coordinate conditions outside the black hole can be affected by the interior data through 
the presence of the mode Eq.~(\ref{coordmode}). The conclusion  is the following: 
the details of how the puncture is handled should not affect the spatial geometry or slicing of
the black hole.  Therefore, 
the boundary conditions at the puncture, whether they are imposed explicitly in one dimension 
or implicitly in three dimensions, do not affect the physics. Their only effect is to 
change the details of how the spatial coordinates are shifted in the spacelike slices. 

Another subtle issue concerning the origin $r=0$ is how the variable $\Gamma^r$ is treated. 
Because $\Gamma^r$ is singular at the puncture, $\Gamma^r \sim -2/r$, one would expect 
the finite difference calculation of 
${\Gamma^r}'$ to generate large errors. This is indeed the case, but the errors are effectively trapped 
near the puncture. I have carried out numerical simulations using the ``regularized" variable 
$\Gamma^r_{reg} \equiv \Gamma^r + 2/r$ in place of $\Gamma^r$; a similar approach is 
advocated in Ref.~\cite{Ruiz:2007rs}. In some cases this does help reduce the errors near the 
puncture, but it does not change the accuracy of the code elsewhere. 

Finally, let me mention that at the outer boundary of the computational domain I use two layers of guard cells 
and simply freeze the data at those points. 
In general the outer boundary is placed at a coordinate distance that is greater than the total run time. 
Assuming causal propagation, the outer boundary conditions do not affect the fields near the 
puncture during the course of a simulation. 

%%%%%%%%%%%%%%%%%%%%%%%%%%%%%%%
\section{Kruskal--Szekeres diagram}
One of the  benefits of assuming spherical symmetry is that we can visualize the motion of the slices and 
grid points on a 
Kruskal--Szekeres diagram. In Kruskal--Szekeres coordinates, the Schwarzschild black hole metric is \cite{Misner:1974qy}
\begin{equation}\label{Kruskalmetric}
  ds^2 = \frac{32 M^3}{R} e^{-R/2M} \left( -dv^2 + du^2 \right) + R^2 \, d\Omega^2  \ .
\end{equation}
Each point in the $u$--$v$ plane is a sphere of areal radius $R$, where $R$ is defined by 
\begin{equation}\label{arealradius}
  u^2 - v^2 = \left( \frac{R}{2M} - 1 \right) e^{R/2M} \ .
\end{equation}
Our goal is  to track the numerical grid points as they move through the $u$--$v$ plane.

There are a number of ways one can relate the numerical data to the $u$--$v$ values of the grid points. I have 
had the most success with the following approach. Consider the metric for a spherically 
symmetric spacetime in 3+1 notation: 
\begin{equation}\label{admmetric}
  ds^2 = -(\alpha^2 - \beta^r\beta_r)dt^2 + 2\beta_r dt dr + \frac{{ g}_{rr}}{\chi} dr^2 + 
  \frac{{ g}_{\theta\theta}}{\chi} d\Omega^2 \ .
\end{equation}
Note that $\beta_r \equiv \beta^r { g}_{rr}/\chi$. Let ${\cal U} = {\rm const}$ be a null curve
in the $r$--$t$ plane. By solving $d{\cal U} \equiv \partial_t{\cal U}\, dt + \partial_r{\cal U}\, dr = 0$
for $dt$ and inserting the result into $ds^2 = 0$, 
we can easily show that ${\cal U}$ satisfies 
${\hat\partial}_\perp {\cal U} = \pm {\hat\partial}_r {\cal U}$. Here, the operator 
${\hat\partial}_\perp \equiv (\partial_t - \beta^r \partial_r)/\alpha$ is the normalized  
derivative orthogonal to the spacelike hypersurface acting on scalars. The operator 
${\hat\partial}_r \equiv \sqrt{\chi/{ g}_{rr}}\partial_r$ is the normalized derivative 
tangent to the spacelike hypersurface. Now observe that  $u\pm v$  are null coordinates. It follows that 
$u$ and $v$ satisfy the equations ${\hat\partial}_\perp (u+ v) = {\hat\partial}_r (u+ v)$ and 
${\hat\partial}_\perp (u- v) = -{\hat\partial}_r (u- v)$, where the
signs are chosen so that $\partial_r u$ is positive. 

The reasoning above shows that  the Kruskal--Szekeres coordinates satisfy the advection equations 
\begin{subequations}
\begin{eqnarray}\label{uvadvection}
  \partial_t (u + v) & = & (\beta^r + \alpha\sqrt{\chi/{ g}_{rr}} ) \partial_r(u + v) \ ,\\
  \partial_t (u - v) & = & (\beta^r - \alpha\sqrt{\chi/{ g}_{rr}} ) \partial_r(u - v) \ .
\end{eqnarray}
\end{subequations}
We can integrate these equations forward in time along with the BSSN variables. To do so we 
must choose initial data and boundary conditions.
The initial data for puncture evolution of a single black hole  is obtained from the Schwarzschild metric 
in isotropic coordinates, 
\begin{subequations}\label{initialdata}
\begin{eqnarray}
  { g}_{rr} & = & 1 \ ,\\
  { g}_{\theta\theta} & = & r^2 \ ,\\
  \chi & = & \left( 1 + {M}/{2r} \right)^{-4} \ ,
\end{eqnarray}
\end{subequations}
along with vanishing extrinsic curvature. 
The areal radius is 
$R \equiv \sqrt{{ g}_{\theta\theta}/\chi} = r(1 + M/2r)^2$. If we assume that the initial slice is
$v=0$,  Eq.~(\ref{arealradius}) shows that $u = F(r)$ where $F(r)$ is defined by 
\begin{equation}
  F(r) \equiv \sqrt{\frac{r}{2M}} \left( 1 - \frac{M}{2r} \right) e^{r (1 + M/2r)^2/(4M)} \ .
\end{equation}
More generally, we can take the initial slice to be boosted in the Kruskal--Szekeres diagram:
\begin{subequations}\label{uvinitialdata}
\begin{eqnarray}
  u & = & F(r)  \cosh\left(t_0/4M\right) \ ,\\
  v & = & F(r)  \sinh\left(t_0/4M\right) \ .
\end{eqnarray}
\end{subequations}
The constant $t_0$ represents the invariance of the Schwarzschild  geometry under the action 
of the Killing vector field $v\partial_u + u\partial_v$.  If we take $t_0 \ne 0$, then the initial data slice 
is related to $v=0$ by sliding the slice along the orbits of the Killing vector field. 

The freedom to choose $t_0$ is useful, because starting from $v=0$ the slice quickly  becomes 
stretched around the singularity $R = 0$. That is, on a Kruskal--Szekeres diagram, 
the slice becomes visually difficult to distinguish from the singularity after a few $M$ of evolution time. 
We can avoid this somewhat by starting 
with an initial slice with $t_0 < 0$. In that case the early time slices are difficult to visualize, 
but the slices at times $t\approx t_0$ are clearly displayed in relation to the singularity and horizon. 
Another useful technique is to slide the slice 
along the Killing vector field after each timestep. This can be used, for example, to keep grid points 
from crossing the $v$ axis, or to fix the location in the Kruskal--Szekeres diagram 
where the slices cross the horizon. 

The variables $u\pm v$ are left or right--moving, depending on the sign of 
$\beta^r \pm \alpha\sqrt{\chi/{ g}_{rr}}$. For 
puncture evolutions,  within a coordinate time of a few $M$, the 
shift vector grows  and the lapse collapses so that  $\beta^r$  dominates over 
$\alpha\sqrt{\chi/{ g}_{rr}}$ near $r=0$.
Therefore, beyond a time of a few $M$,  both characteristic fields are outgoing (toward decreasing $r$) at the boundary $r=0$. 
In the vicinity of the origin I use 
one--sided differencing for $u$ and $v$, appropriate for outgoing fields. Thus, no guard cells are needed 
for $u$ and $v$ at $r=0$. 

At the outer boundary of the computational domain we have  $\beta^r \approx 0$ and 
$\alpha\sqrt{\chi/{ g}_{rr}} \approx 1$. Thus,  the characteristic field $u+v$ is incoming and the field $u-v$ is 
outgoing. Boundary conditions are applied with the following scheme. 
First, let us equate the coefficients of $dt^2$ in Eqs.~(\ref{Kruskalmetric}) and (\ref{admmetric}) to obtain
\begin{equation}\label{coefficientofdt2}
  {\dot u}^2 - {\dot v}^2 = -\frac{R}{32M} e^{R/2M} (\alpha^2 - \beta^r \beta_r) \ ,
\end{equation}
where the dot denotes $\partial_t$. Now differentiate the definition (\ref{arealradius}) with respect to 
time to obtain a second equation involving $\dot u$ and $\dot v$. Together these equations have the solution
\begin{subequations}\label{uvodes}
\begin{eqnarray}
  {\dot u} & = & \frac{1}{2G} \left( u {\dot G} 
      + \varepsilon v \sqrt{{\dot G}^2 + 4GH(\alpha^2 - \beta^r\beta_r)} \right) \ ,\quad  \\
  {\dot v} & = & \frac{1}{2G} \left( v {\dot G} 
      + \varepsilon u \sqrt{{\dot G}^2 + 4GH(\alpha^2 - \beta^r\beta_r)} \right) \ ,\quad 
\end{eqnarray}
\end{subequations}
where $G \equiv (R/2M - 1)\exp(R/2M)$, $H \equiv (R/32M^3)\exp(R/2M)$, and $\varepsilon \equiv u/|u|$. 
Note that ${\dot G}$ can be written in terms of $R$ and $\dot R$. Also recall that the areal radius can be written 
in terms of the BSSN variables as $R = \sqrt{{ g}_{\theta\theta}/\chi}$, so the time derivative $\dot R$ depends 
on $\partial_t{ g}_{\theta\theta}$ 
and $\partial_t\chi$. The time derivatives of these BSSN variables can be replaced with the right--hand sides of the BSSN
equations (\ref{chieqn}) and (\ref{gthetathetaeqn}). 

Equations (\ref{uvodes}) are a set of ordinary differential equations for $u$ and $v$. These can be integrated forward 
in time without the calculation of any spatial derivatives.  I use these equations to determine $u$ and $v$ in the 
vicinity of  the outer boundary, providing a layer of two guard cells for the interior calculation based on 
Eqs.~(\ref{uvadvection}). 

One can try to use Eqs.~(\ref{uvodes}) to evolve $u$ and $v$ everywhere.  
In practice this does not work well. It is difficult to 
integrate these equations across the horizons where $G = 0$. Also it is difficult to maintain accuracy and stability as 
grid points evolve across the $v$ axis, where the value of $\varepsilon$ changes from $-1$ to $+1$. 

Finally, let me mention a method for placing the data of a single time slice 
into a Kruskal--Szekeres diagram. From the numerical data one can compute the areal radius $R$ and the proper 
distance from the horizon, $L$, for each grid point. In the $u$--$v$ plane the metric gives
\begin{equation}
	\Delta L^2 = \frac{32M^3}{R} e^{-R/2M} (\Delta u^2 - \Delta v^2) \ .
\end{equation}
We also find, by differentiating the definition (\ref{arealradius}), 
\begin{equation}
	u \Delta u - v\Delta v = \frac{R}{8M^2} e^{R/2M} \Delta R \ .
\end{equation}
These equations can be solved for $\Delta u$ and $\Delta v$ as functions of $\Delta R$ and $\Delta L$. 
The values of $u$ and $v$ for each grid point are found by numerical integration, starting from 
some initial values for $u$ and $v$. These initial values must 
satisfy the restriction (\ref{arealradius}), which leaves some freedom of choice. For example, 
if the intitial areal radius is less than $2M$, one can choose the initial values to be $u=0$ 
and $v = \sqrt{1 - R/2M}e^{R/4M}$. 
The freedom in choosing the initial values for $u$ and $v$ is a consequence of the  
Killing isometry of Schwarzschild spacetime. 

%%%%%%%%%%%%%%%%%%%%%%%%%%%%%%%%%%
\section{Results}
There are a large number of cases that we can consider, depending on our choice of Eulerian versus Lagrangian 
evolution for $g$, the advection terms that we include or omit in the slicing and 
coordinate conditions, and the value for the damping parameter $\eta$ that appears in the 
$\Gamma$--driver shift. 
I will not attempt to provide a complete examination and comparison of all these possibilities here. Rather, I 
will give a sampling of some of the interesting results. 

The initial metric for puncture evolution is given in Eq.~(\ref{initialdata}). 
Initially, the conformal connection function 
is ${\Gamma}^r = -2/r$, and the extrinsic curvature components ${A}_{rr}$ and $K$ vanish. The shift 
vector $\beta^r$ and the auxiliary field $B^r$ are chosen to vanish at the initial time. The initial lapse function 
is either unity, $\alpha = 1$, or has the ``pre--collapsed'' form  $\alpha = (1 + M/2r)^{-2}$. 
In all cases the code uses fourth order finite differencing in space and fourth order Runge--Kutta for time 
stepping.

The geometrical description of puncture evolution for single black holes is now well understood 
\cite{Hannam:2006vv,Hannam:2006xw,Brown:2007tb,Garfinkle:2007yt,Hannam:2008sg}. The 
initial geometry is a wormhole. 
As the evolution begins, the grid points near the ``other" asymptotically flat end quickly shift 
into the black hole interior. The numerical data then settle onto a portion of a ``trumpet 
slice" of the black hole. Such a slice asymptotes to a constant areal radius of about $1.3M$. The key 
ingredient responsible for this behavior is the $\Gamma$--driver shift condition. Tests using 
the one--dimensional code with $\Gamma$--driver shift invariably show the same qualitative 
evolution for the physical geometry, the same evolution from 
wormhole to trumpet, for all of the cases $E_\pm^+$ and $L_\pm^\pm$. However, the detailed 
behavior of the GBSSN variables can differ significantly from case to case. 

\begin{figure}[tb!]
\includegraphics[scale=0.9, viewport=135 350 440 540]{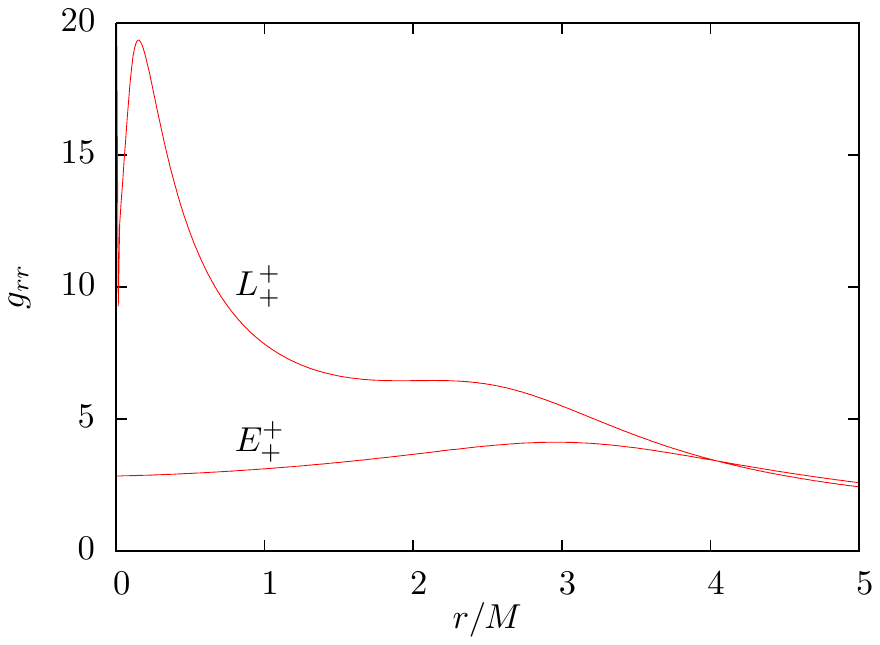}
\caption{The conformal metric component $g_{rr}$ near the puncture $r=0$ 
at time $t=50M$ for the systems $E^+_+$ and $L^+_+$.}
\label{figure2compA}
\end{figure}

\begin{figure}[t!]
\includegraphics[scale=0.9, viewport=135 350 440 540]{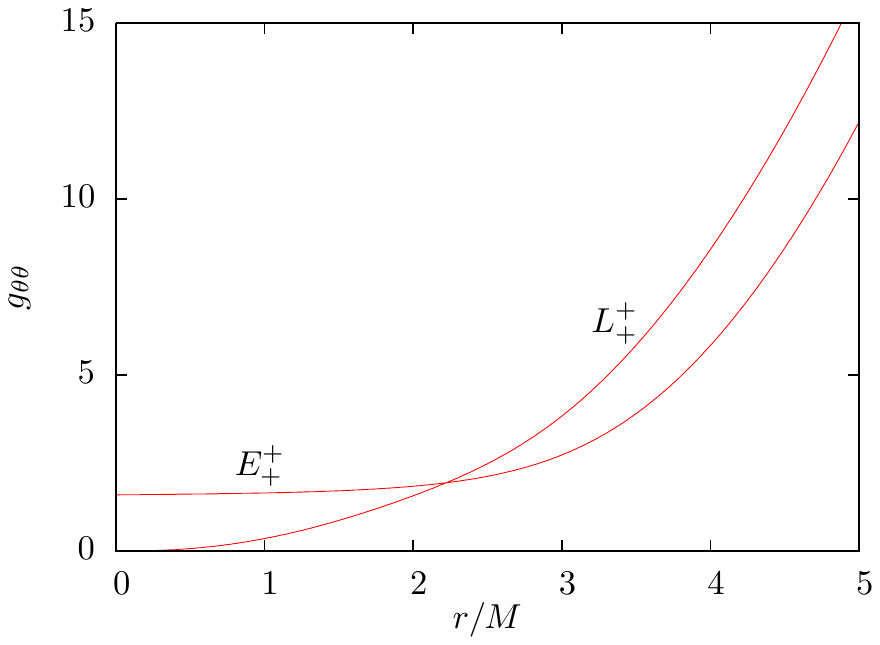}
\caption{The conformal metric comonent $g_{\theta\theta}$ at $t=50M$ for $E^+_+$ and $L^+_+$.}
\label{figure2compB}
\end{figure}

\begin{figure}[t!]
\includegraphics[scale=0.9, viewport=135 350 440 540]{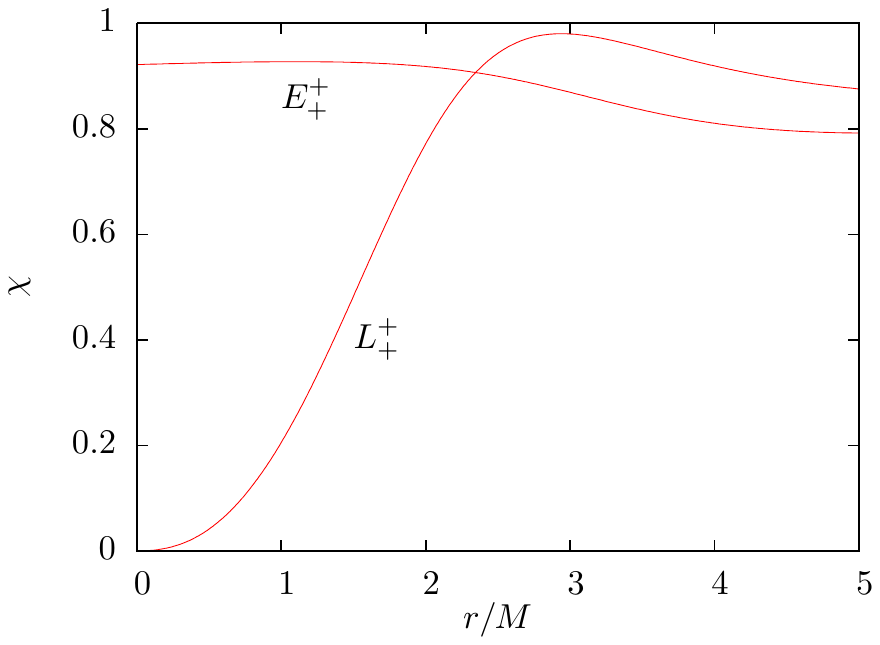}
\caption{The conformal factor $\chi$ at $t=50M$ for $E^+_+$ and $L^+_+$.}
\label{figure2compC}
\end{figure}

\begin{figure}[t!]
\includegraphics[scale=0.9, viewport=135 350 440 540]{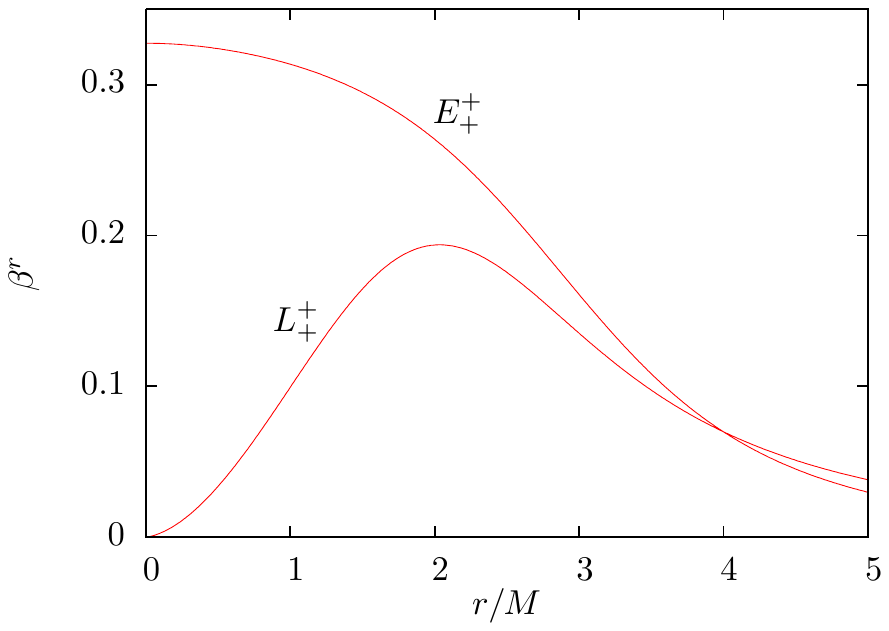}
\caption{The shift vector $\beta^r$ at $t=50M$ for $E^+_+$ and $L^+_+$.}
\label{figure2compD}
\end{figure}

Figures (\ref{figure2compA}--\ref{figure2compD}) show the metric 
components $g_{rr}$, $g_{\theta\theta}$, the conformal factor $\chi$, and the shift vector $\beta^r$ at 
time $t = 50M$ for the cases $E_+^+$ and $L_+^+$. The data for these graphs were taken from simulations 
with resolution $\Delta r = M/100$. 
The differences near the puncture are evident. 

One might guess that the difference between the cases $E^+_+$ and $L_+^+$ is a 
consequence of the different boundary conditions imposed at $r=0$. 
However, the Eulerian equations cannot be evolved with the Lagrangian boundary conditions (\ref{LagrangianBC}). 
Recall that the Lagrangian boundary conditions are designed to keep  
$g_{\theta\theta}$ and $\beta^r$ equal to zero at the puncture. If these conditions are used 
with the Eulerian evolution equations, the values of $g_{\theta\theta}$ and $\beta^r$ near the 
puncture grow rapidly and a sharp gradient develops between the first few grid points $j=1,2,\dots$ and the guard 
cell $j=0$. The code crashes shortly after $t=1M$. With the boundary conditions (\ref{EulerianBC}), the 
fields $g_{\theta\theta}$ and $\beta^r$ are allowed to develop nonzero values at the puncture. Figures 
(\ref{figure2compB}) and (\ref{figure2compD}) show that indeed $g_{\theta\theta}$ and $\beta^r$ evolve to nonzero 
values in the Eulerian case $E_+^+$. 

The converse is also true: The Lagrangian equations will not evolve stably 
with the Eulerian boundary conditions (\ref{EulerianBC}). 
If one attempts such an evolution, the code quickly develops problems at the puncture and crashes. 
It appears that the behavior of the fields near the puncture is primarily dictated by the 
equations of motion in the bulk.  Perhaps the main role played by the 
boundary conditions (\ref{LagrangianBC}) or (\ref{EulerianBC}) is to help insure numerical stability 
for the GBSSN equations in spherical symmetry. 

Although the results obtained with $E^+_+$ and $L^+_+$ appear quite different, they represent the same 
slicing of the same spacetime geometry. It is clear that the spacetime geometry is that of a 
Schwarzschild black hole, since a Schwarzschild black hole is the unique solution of the 
vacuum Einstein equations with the chosen initial data. It is not quite so obvious that the 
evolutions $E^+_+$ and $L^+_+$ lead to the same foliation of that spacetime geometry. The difference between 
the Eulerian and Lagrangian cases resides in the choice for $\dperp\ln g$, as seen in 
Eqs.~(\ref{bssnequations}). The terms containing $\dperp\ln g$ can be absorbed into redefinitions 
of the field variables. For example, Eq.~(\ref{bssnequations}a) can be written as 
$\dperp \hat\phi = -\alpha K/6$ where $\hat\phi \equiv \phi + \ln g^{1/12}$. 
Similarly, Eq.~(\ref{bssnequations}b) can be 
written as $\dperp \hat g_{ab} = -2\alpha A_{ab}$ where $\hat g_{ab} \equiv g^{-1/3}g_{ab}$. 
The physical metric (\ref{physicalgandK}a) is invariant under this redefinition: 
$e^{4\hat \phi} \hat g_{ab} = e^{4\phi} g_{ab}$. This invariance is a consequence of the conformal 
symmetry of the GBSSN equations \cite{Brown:2005aq}. In this way we see that, independent of the choice for 
$\dperp\ln g$, the physical metric and extrinsic curvature undergo identical evolutionary paths. 

\begin{figure}[t!]
\includegraphics[scale=0.9, viewport=135 350 440 540]{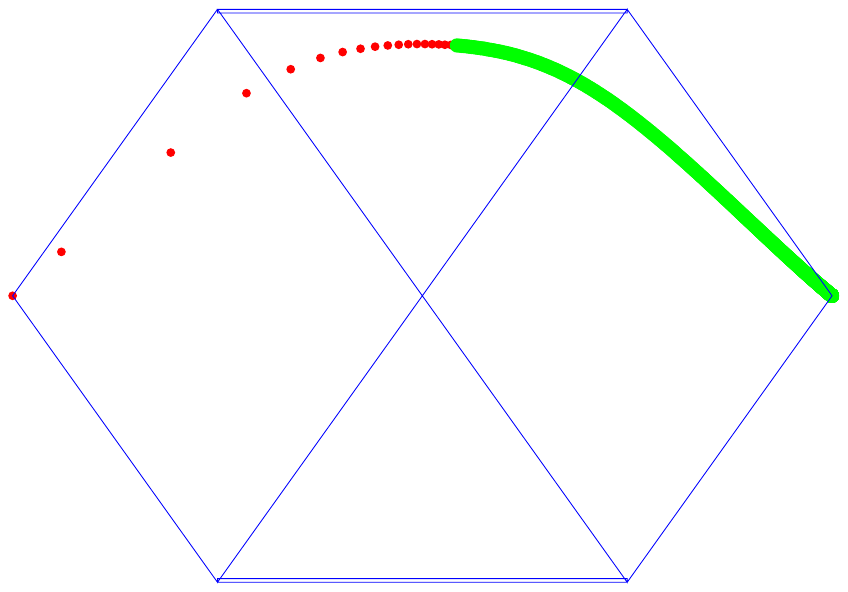}
\caption{Penrose diagram showing the grid points for the cases $E^+_+$ (large green circles) 
and $L^+_+$ (small red circles) at time $t=2.5M$. The data for $E^+_+$ blend together to form a thick 
curve that ends in the middle of the diagram.}
\label{figure2penrose}
\end{figure}

Figure (\ref{figure2penrose}) shows a penrose diagram with the grid points at time $t = 2.5M$ as obtained 
from the $E^+_+$ and $L_+^+$ cases. 
The slice is the same for the two cases. What is clearly different is the coverage  of that slice by 
the available grid points. In the Eulerian case, the relatively large positive shift near the puncture, 
seen in Fig.~(\ref{figure2compD}), drives the grid points away from the left spacelike infinity and 
into the interior of the black hole. In the Lagrangian case the grid points are also driven from 
the left spacelike infinity into the black hole interior, but not as quickly. 

\begin{figure}[t!]
\includegraphics[scale=0.9, viewport=135 350 440 540]{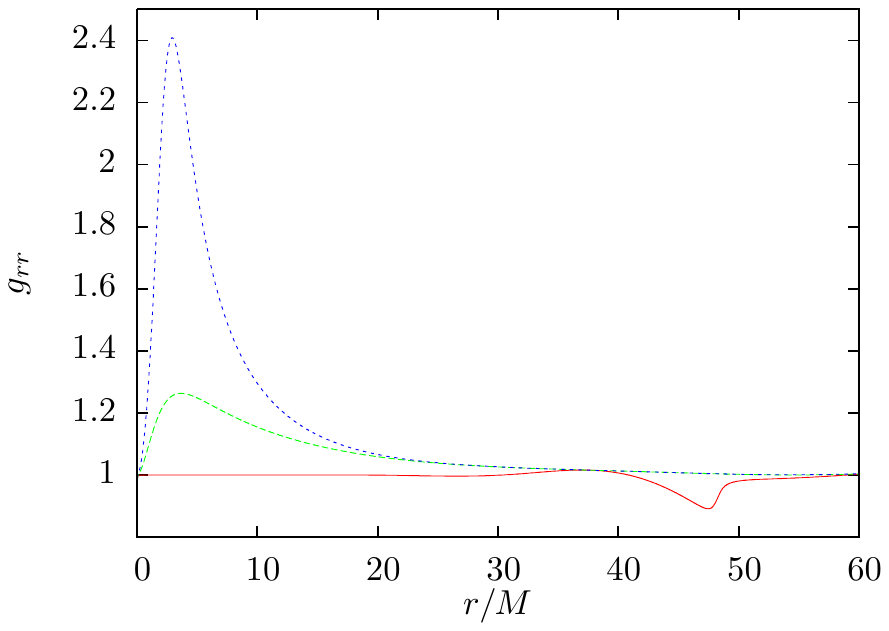}
\caption{${g}_{rr}$ as a function of coordinate radius $r$ at time $t=50M$ 
for the case $L^+_-$.  The three curves 
correspond to values $0$, $1/M$, and $2/M$ for the damping parameter $\eta$. The 
peak values increase with increasing $\eta$.}
\label{figure21a}
\end{figure}

\begin{figure}[t!]
\includegraphics[scale=0.9, viewport=135 350 440 540]{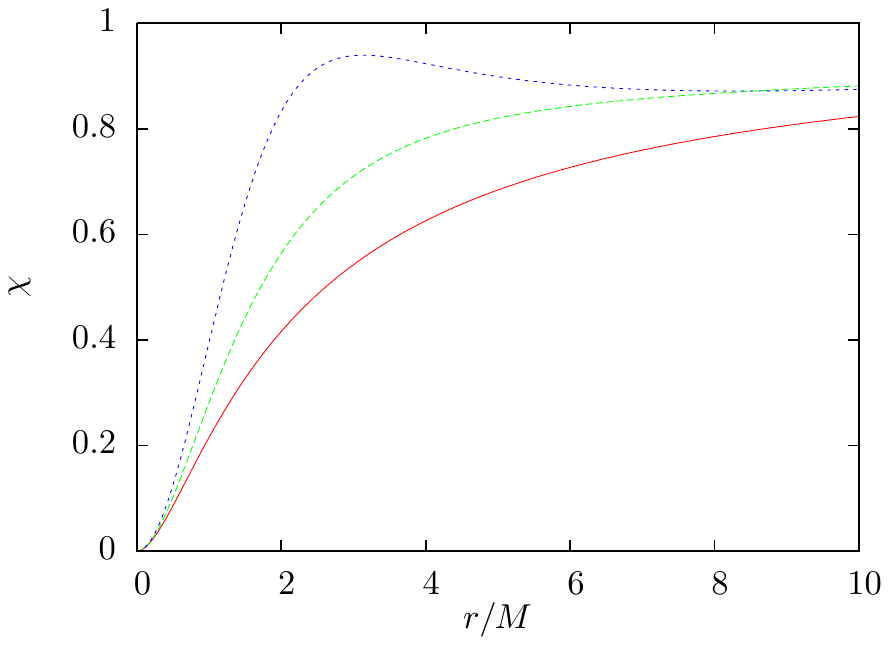}
\caption{$\chi$ as a function of  $r$ at time $t=50M$ 
for the case $L^+_-$ with $\eta = 0$, $1/M$, and $2/M$. The 
peak values increase with increasing $\eta$.}
\label{figure21b}
\end{figure}

\begin{figure}[t!]
\includegraphics[scale=0.9, viewport=135 350 440 540]{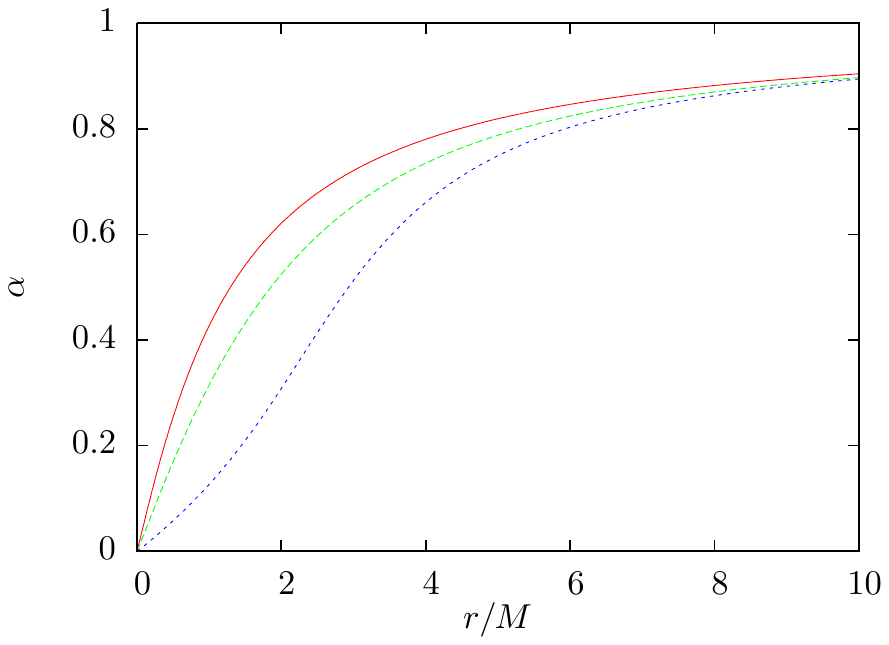}
\caption{$\alpha$ as a function of  $r$ at time $t=50M$ 
for the case $L^+_-$ with $\eta = 0$, $1/M$, and $2/M$. The 
peak values decrease with increasing $\eta$.}
\label{figure21c}
\end{figure}

\begin{figure}[t!]
\includegraphics[scale=0.9, viewport=135 350 440 540]{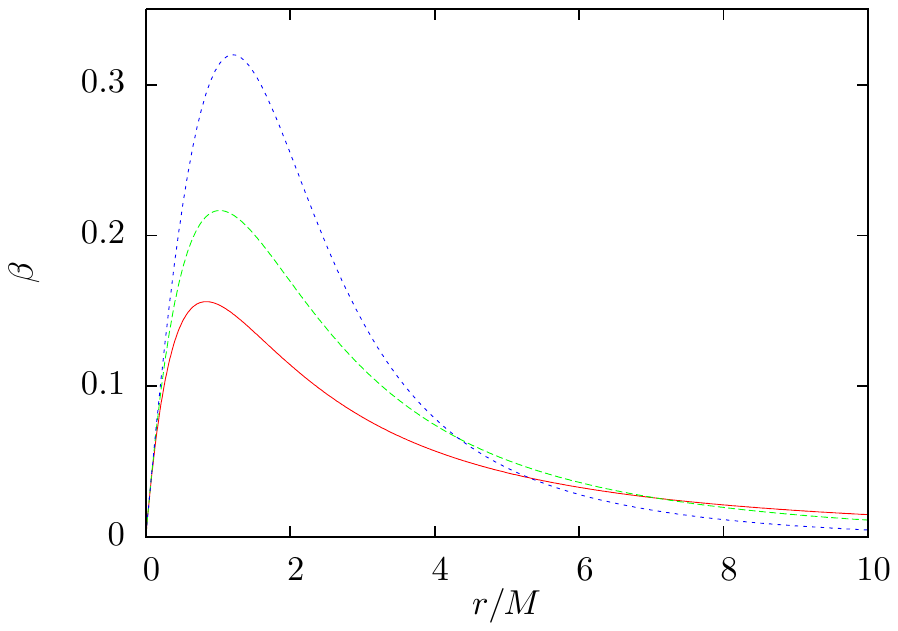}
\caption{$\beta^r$ as a function of  $r$ at time $t=50M$ 
for the case $L^+_-$ with $\eta = 0$, $1/M$, and $2/M$. The 
peak values increase with increasing $\eta$.}
\label{figure21d}
\end{figure}

The behavior of the three--dimensional GBSSN equations in the case $L^+_-$ appears to be close to 
the behavior of three-dimensional BSSN codes. It is particularly useful to compare the  
results presented below with those of Ref.~\cite{Brugmann:2008zz}. 
In particular, Figs.~(\ref{figure21a}--\ref{figure21d}) show graphs of 
the variables ${g}_{rr}$, $\chi$, $\alpha$, and $\beta^r$ as functions of the 
coordinate radius $r$ at time $t=50M$. The evolution is type $L^+_-$ (Lagrangian case with 
1+log slicing including the advection 
term and $\Gamma$--driver shift excluding advection terms), and begun with a pre--collapsed 
lapse. The resolution used for these simulations was $\Delta r = M/100$. 
The curves  in these figures show the results obtained with three values of the damping parameter,  
$\eta=0$, $\eta=1/M$, and $\eta=2/M$. 

Recall that $\eta$ appears in the $\Gamma$--driver shift condition. 
For 1+log slicing including the advection term the shift vector does not affect the slicing 
of spacetime or the spatial geometry, it only 
changes the coordinate system (or distribution of grid points) from one slice to the next \cite{Brown:2007tb}.  
Thus the results obtained with different values of $\eta$ are physically equivalent. Nevertheless, 
there is reason to prefer 
a value that will allow the lapse and shift to adjust to the Killing symmetry of the Schwarzschild 
geometry. That is, we want the gauge to be ``symmetry seeking'' and bring the data close to a stationary 
state at late times \cite{Garfinkle:2007yt}. For the one--dimensional code with 1+log 
slicing the value $\eta=0$ is a good choice. 
With $\eta=0$ the data settle very close to a stationary state by $t=50M$. 

With positive $\eta$ the data continue to evolve indefinitely. This tendency has been seen in three--dimensional 
simulations \cite{Brugmann:2008zz} where it is described as ``coordinate drift''. We can monitor this drift by computing the L2 norm of the right--hand sides of the GBSSN equations. (The L2 norm is 
defined as a sum over the right--hand sides 
of the evolution equations for ${g}_{rr}$, $\chi$, $K$, ${\tilde \Gamma}^r$, $\alpha$, and $\beta^r$, and 
a sum over grid points between $r=0$ and $r=25M$.) 
\begin{figure}[t!]
\includegraphics[scale=0.9, viewport=135 350 440 540]{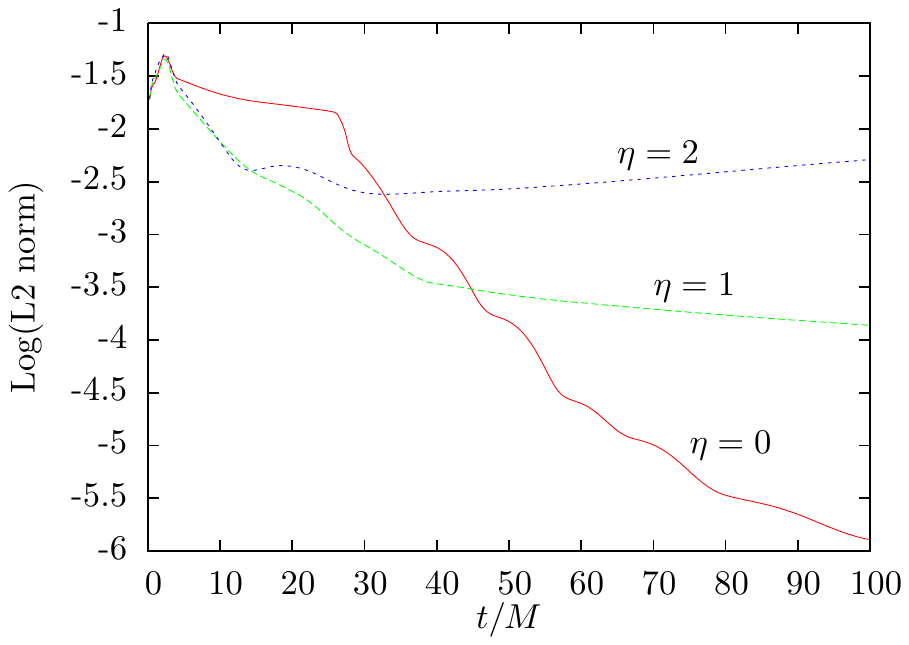}
\caption{Common Logarithm of the L2 norm of the right--hand sides of the equations 
of motion versus time.}
\label{figure20}
\end{figure}
For each value of $\eta$ the L2 norm rises to a peak value just beyond $t = 2M$. For $\eta=0$ 
the norm drops slowly between $t\approx 2M$ and $t\approx 25$, then drops rapidly for $t>25M$. The sudden 
change in slope at $t\approx 25M$ occurs for the following reason. At early times there 
is a relatively large adjustment in the 
geometry near the puncture as the grid points near the puncture pull away from spacelike infinity, 
enter the black hole interior, and  relax to the ``trumpet slice" \cite{Brown:2007tb}. This adjustment 
can be seen in the values of the right--hand sides as a pulse that propagates outward. 
At $t\approx 25M$ the pulse passes beyond the region 
in which the L2 norm is computed. The rapid decay for $\eta=0$ beyond $t\approx 25M$ shows that 
the data in the interior region $r<25M$ of the computational grid is very effectively becoming stationary. 

The code described here uses fourth order finite differencing in space and time. 
However, the code does not always exhibit fourth order convergence near the puncture. 
\begin{figure}[t!]
\includegraphics[scale=0.9, viewport=135 350 440 540]{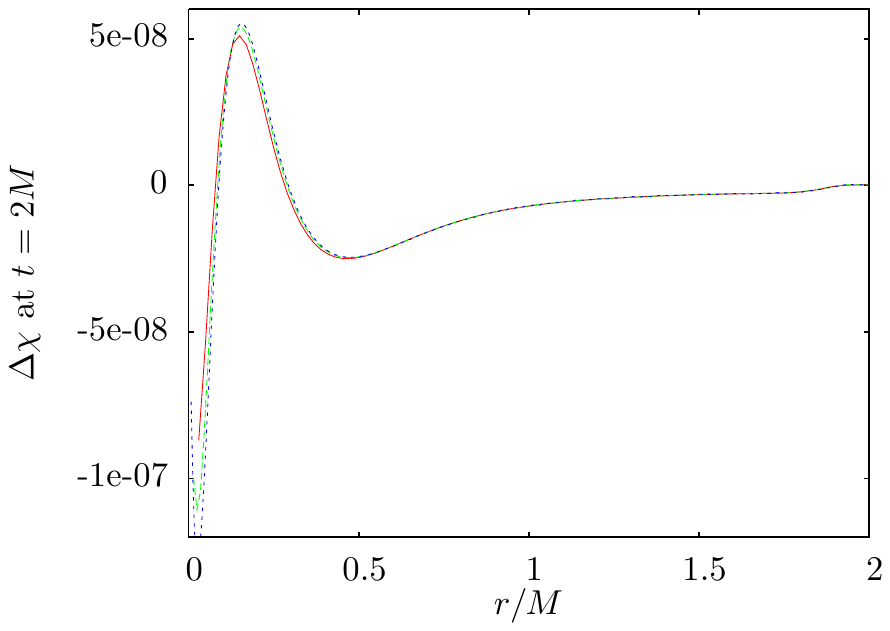}
\caption{Convergence plot for $\chi$ at $t=2M$. The data are obtained from 
simulations at resolutions $\Delta r = M/50$, $M/100$, $M/200$, and $M/400$. The 
curves are scaled by powers of $16$, appropriate for fourth--order convergence.}
\label{figure22a}
\end{figure}
\begin{figure}[t!]
\includegraphics[scale=0.9, viewport=135 350 440 540]{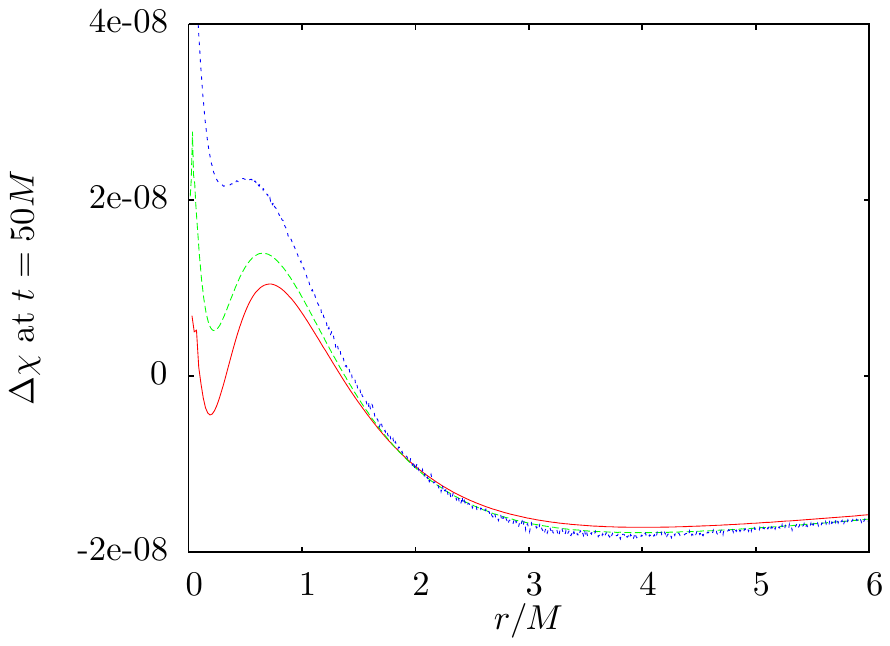}
\caption{Convergence plots for $\chi$ at $t=50M$. The resolutions and scaling are 
the same as those used in Fig.~(\ref{figure22a}).}
\label{figure22b}
\end{figure}
Figures (\ref{figure22a}) and (\ref{figure22b}) show
convergence plots for $\chi$ at early and late times, $t=2M$ and $t=50M$, for the case $L^+_-$. 
Each graph shows three curves, obtained 
by computing differences between values of $\chi$ at successive resolutions and scaling by powers of $16$. 
To be explicit, let us add a subscript to $\chi$ to 
denote the resolution $\Delta r$. Then the curves in Figs.~(\ref{figure22a}) and (\ref{figure22b}) are 
$(\chi_{M/50} - \chi_{M/100})$, $16(\chi_{M/100} - \chi_{M/200})$, and $256(\chi_{M/200} - \chi_{M/400})$.
These graphs show that
at early times the data are fourth--order convergent, while at late 
times the data are fourth order convergent outside the 
region close to the puncture. 

\begin{figure}[t!]
\includegraphics[scale=0.9, viewport=135 350 440 540]{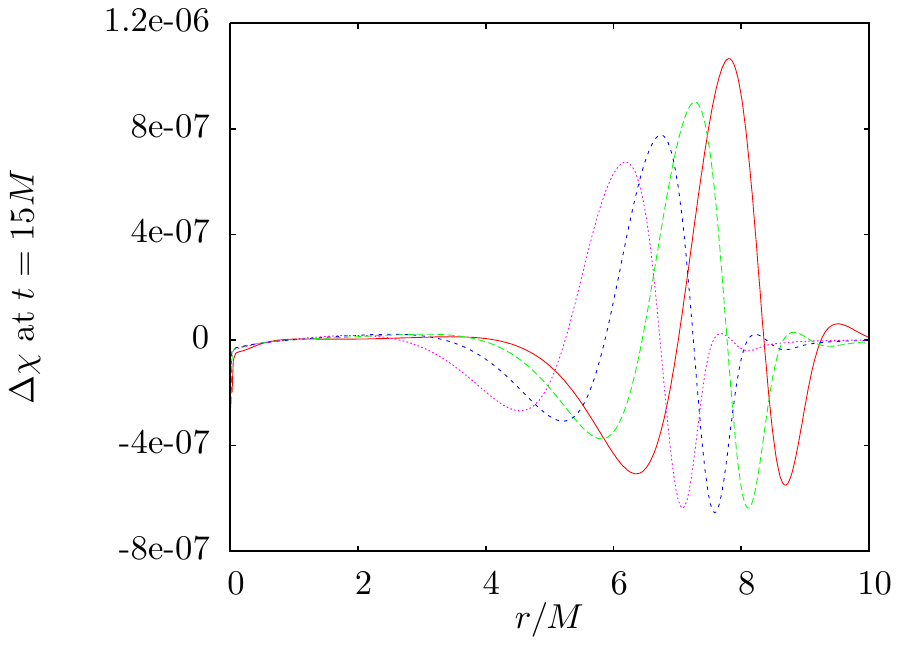}
\caption{Convergence plot for $\chi$ at $t=15M$. The resolutions are the same 
as those used in the previous figures. In this case the 
curves are scaled by powers of $4$, appropriate for second--order convergence.}
\label{figure23}
\end{figure}

Figure (\ref{figure23}) is a convergence plot for $\chi$ at the intermediate time $t=15M$, again 
for the system $L^+_-$. 
This graph shows four curves, $(\chi_{M/50} - \chi_{M/100})$, $4(\chi_{M/100} - \chi_{M/200})$, $16(\chi_{M/200} - \chi_{M/400})$,
and $64(\chi_{M/400} - \chi_{M/800})$. The curves are scaled by powers of $4$, which is the 
scaling expected for second--order convergence. An error pulse is generated near the puncture and propagates  
outward through the computational domain. Note that this error is delayed at higher resolution, 
resulting in a shift in the curves. 

These tests suggest that the puncture generates an approximately second--order error of the form
\begin{equation}
 \chi_{\Delta r} = \chi_{\rm exact} + \frac{\Delta r^2}{r}{\cal E}(r - t + 2M\log(M/\Delta r)) 
\end{equation}
for some function ${\cal E}$. The pulse appears to form from errors near the puncture. Note that the 
horizon begins at coordinate radius 
$r=M/2$ and stays within $r \leq M$ throughout the evolution.  The pulse moves with a speed 
of approximately $1$. 
It has an amplitude approximately proportional to $\Delta r^2/r$. At higher resolution the pulse is delayed. 
This delay appears as a shift toward the puncture in Fig.~(\ref{figure23}). 
The amount of delay or shift  is  proportional to the 
logarithm of $1/\Delta r$. 

Similar convergence tests with the $E^+_+$ system do not show this anomolous behavior. This suggests 
that the origin of the non--fourth order error for $L^+_-$ is related to the presence of  
modes with the unusual proper speed $\hat\beta^r$. Because the 
characteristic fields and speed for traditional BSSN are more closely related to those for $E_+^+$,
I would not expect to see such behavior in current 3D codes using the traditional BSSN equations. 

\begin{figure}[b!]
\includegraphics[scale=0.85, viewport=150 350 400 520]{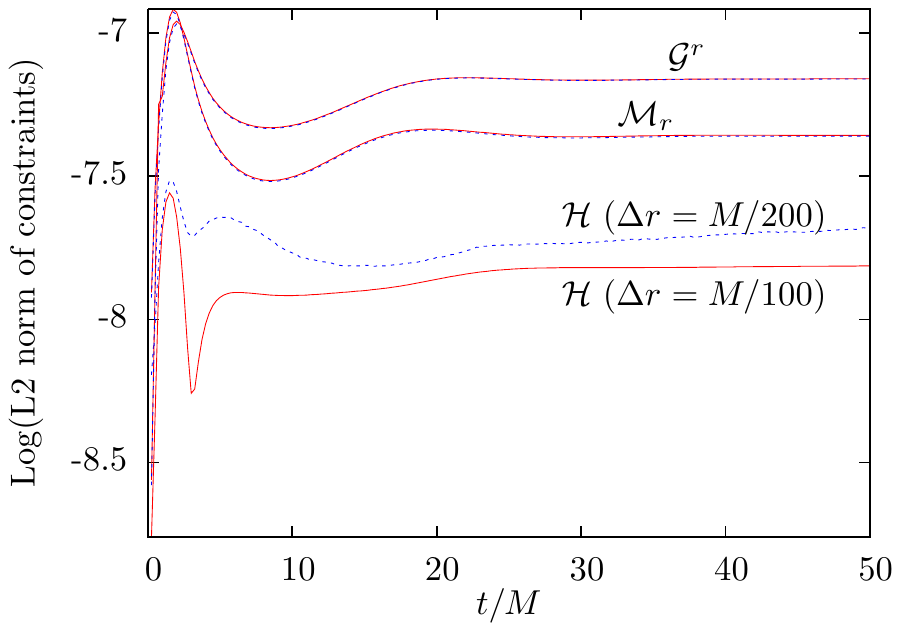}
\caption{Common log of the L2 norm of the constraints ${\cal H}$, ${\cal M}_r$, and ${\cal G}^r$. For 
each constraint two resolutions are plotted, $\Delta r = M/100$ and $\Delta r = M/200$. For the 
higher resolution, the constraints are multiplied by a factor of $16$. The two curves for ${\cal M}_r$
overlap one another, and are nearly indistinguishable in this figure. Likewise the two curves 
for ${\cal G}^r$ overlap one another.}
\label{figureconstraints}
\end{figure}

Figure (\ref{figureconstraints}) shows a plot of the L2 norm of the constraints (\ref{constraints}) 
as functions of time. The constraint values are stable over time, with only the Hamiltonian
constraint ${\cal H}$ showing a slight upward drift. 
All of the constraints exhibit large non--convergent errors near the puncture. These errors dominate the 
calculation of the L2 norm over the entire computational domain. As a result, the data for 
Fig.~(\ref{figureconstraints}) was obtained by omitting the region $0<r<0.5M$. The data 
beyond $r>25M$ is also excluded so that outer boundary effects are ignored. The 
data for the momentum constraint ${\cal M}_r$ and the conformal connection function constraint 
${\cal G}^r$ show nearly perfect fourth order convergence over this region $M/2 < r < 25M$. 
The results for the Hamiltonian constraint ${\cal H}$ are not so ideal. I suspect this is related 
to errors that arise in computing ${\cal H}$, which depends on second spatial derivatives 
of the metric components $g_{\theta\theta}$ and $\chi$. The constraints ${\cal M}_r$ and ${\cal G}^r$ 
do not involve second order derivatives. 

\begin{figure}[b!]
\includegraphics[scale=0.85, viewport=150 310 400 470]{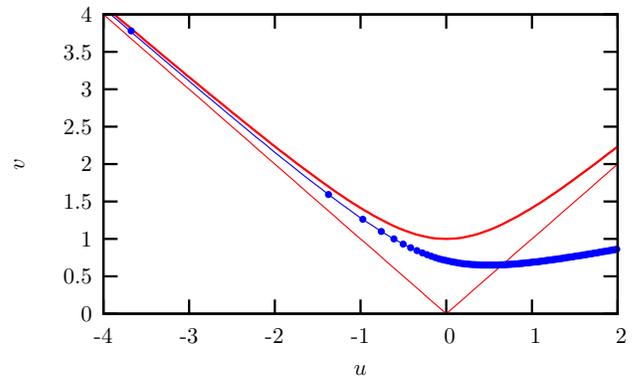}
\caption{Kruskal diagram at $t=50M$ for simulations with 1+log slicing. The hyperbolic curve is the future 
singularity and the lines at $\pm 45^\circ$ are the horizons.}
\label{figure24}
\end{figure}
The flow of grid points on a Kruskal--Szekeres diagram can be particularly enlightening. Figure 
(\ref{figure24}) shows the grid points at $t=50M$ located on a portion of the Kruskal--Szekeres diagram. This data was 
obtained from a simulation $L^+_-$ with $\eta=0$, 
and initial lapse $\alpha = 1$.  
The initial slice in the diagram was obtained from  Eq.~(\ref{uvinitialdata}) with $t_0 = -50M$. 
The resolution for this simulation was $\Delta r = M/100$. For clarity, only the odd numbered grid points are shown in 
this figure. The smooth curve was obtained from a simulation with vanishing shift vector, case $L_0$. 

The first grid point in Fig.~(\ref{figure24}), $j=1$, is at location $u\approx -3.7$, $v\approx 3.7$ 
between the horizon and the physical singularity. 
As shown  in Ref.~\cite{Brown:2007tb}, the grid points near the puncture are 
rapidly drawn into the black hole interior by the 
$\Gamma$--driver shift condition. This is, however, a resolution--dependent effect. 
In the limit of infinitely high resolution, 
the grid points merge into the smooth curve. That curve crosses the horizon and 
asymptotes to spacelike infinity ($u\sim -\infty$, 
$v\sim {\rm finite}$). 

\begin{figure}[t!]
\includegraphics[scale=0.85, viewport=150 310 400 470]{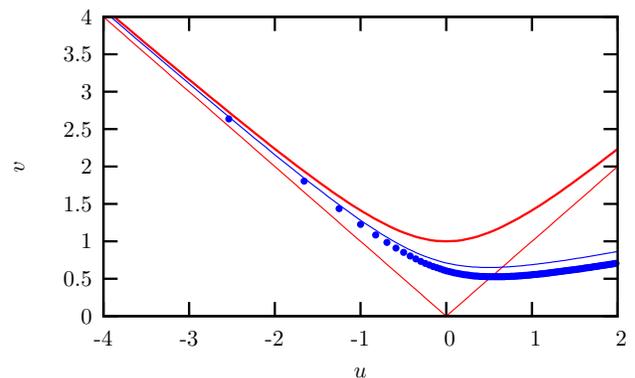}
\caption{Kruskal--Szekeres diagram at $t=50M$ for simulations with slicing condition 
$\partial_t\alpha = -2\alpha K$.}
\label{figure25}
\end{figure}
Figure (\ref{figure25}) compares the results of simulations with cases $L_-^-$ and $L_0$.
Again, we use  $\eta=0$,  unit 
initial lapse, and choose $t_0 = -50M$. The figure shows grid points $j=3$, $5$, $7\ldots$ at time $t=50M$ for the 
$L^-_-$ evolution.  
The grid point $j=1$, which is not shown, is at location $u\approx -12.7$, $v\approx 12.7$. 
The smooth curve on the Kruskal--Szekeres diagram was obtained from the $L_0$ evolution. Note that 
in this case the grid points do not lie on the curve. This is because without the advection term
in the 1+log slicing condition, the slicing  depends on the shift vector. 

Finally, let me comment on the results of long term evolution. Typically, after a run time on the order of 
a few light crossing times, the code will crash due to the development of a shock. For example, with 
$E_+^+$ evolution on a grid with outer boundary at $125M$, steep gradients and spikes develop 
in the BSSN variables at about $r\approx 8.5M$. The behavior causes the code 
to crash at $t\approx 250M$. With the outer boundary placed at $75M$, the shock develops at $r\approx 6M$ 
and the code crashes at $t\approx 160M$. It is possible that improvements to the outer boundary conditions might 
postpone or even eliminate these shocks. On the other hand, it has been suggested that 
gauge shocks are a generic result of 1+log type slicing conditions \cite{Garfinkle:2007yt}. Currently, such 
behavior has only been seen with 1+1 codes. 

%%%%%%%%%%%%%%%%%%%%%%%%%%%%%%%%%%%%%%%%%%%%%%%%%%%%%%%%%%%
\begin{acknowledgments}
I would like to thank Dae-Il Choi, Pablo Laguna, Olivier Sarbach and Manuel Tiglio for helpful discussions. 
This work was supported by NSF grant PHY--0600402. 
\end{acknowledgments}

\bibliography{references}

\end{document}